# Irradiation Tests for Commercial Off-the Shelf Components with Atmospheric-like Neutrons and Heavy-Ions

Paolo Branchini[†], Andrea Fabbri[†] (CA), Sacha Cormenier[†+], Marco Bernardini[†+]

Giovanni Romanelli[††], Enrico Preziosi[††] (CA), Roberto Senesi[††], Carla Andreani[††]

Chris. Frost*, Carlo Cazzaniga*

Toni Fabio Catalano, Mario Buffardo [†††]

[†]INFN, Roma Tre Section, Rome, Italy

+ Maths and Physics Department, Università degli Studi di Roma "Roma Tre", Via della Vasca Navale 84, 00146, Rome, Italy

[††] NAST Centre and Physics Department, Università degli Studi di Roma "Tor Vergata", Via della Ricerca Scientifica 1, 00133, Rome, Italy

[†††] Thales Alenia Space Italia, Rome, Italy

* ISIS Facility, STFC Rutherford Appleton Laboratory, Chilton, Didcot, Oxfordshire OX11 0QX, UK

paolo.branchini@roma3.infn.it, andrea.fabbri@roma3.infn.it, sacha.cormenier@uniroma3.it, marco.bernardini@roma3.infn.it, toni.catalano@thalesaleniaspace.com, mario.buffardo@thalesaleniaspace.com

*Abstract*—**This paper presents the results of the irradiation, performed with atmospheric-like neutrons and heavy-ions, of Commercial Off-the Shelf Components (COTS), which can be used in space missions. In such cases, it is crucial to perform tests in a radiation environment that emulates the environment of different orbits around Earth. In our study we used atmospheric-like neutrons with fluences up to $10^{11}$ neutrons/cm$^{-2}$ and Kr ions of fluences up to $10^7$ ions/cm$^{-2}$. These intensities are augmented with respect to the atmospheric one in order to shorten the irradiation time while simulating a long-time exposure during a possible mission in Low Earth Orbit (LEO). A similar radiation environment to LEO can also be present during High-Energy Physics experiments. Therefore, the study herby reported can also be helpful for accelerator physics. In this paper we show in detail procedures, setup and results we have obtained on a commercial device normally exploited in automotive environments.**

*Index Terms*—**CMOS technology, COTS, electronics, heavy ion, microcontrollers, neutrons, irradiation, SEE, SEL.**

## I. INTRODUCTION

In the present time, electronic devices represent a fundamental part of our lives and they are the main elements for high energy physics experiments, medical applications and space missions, just to name a few. Some of these fields of study take place in environments subjected to various amounts of radiation that can compromise the electronics and the people involved; it is then of paramount importance to study the radiation environment and the effect that such radiations have both on electronics, with a particular interest for Commercial Off-the Shelf (COTS) components, and humans.

COTS components are beginning to be widely used for space missions due to their reduced price and short procurement time. Therefore, a thorough radiation hardness study must be performed on them since the space environment presents radiation hazards from particle fluxes that can produce effects like:

- Single Event Effect (SEE),
- latch-up,
- erosion,
- Total Ionizing Dose effect (TID),
- displacement,
- charging,
- interference.

As reported in [1] there are already some devices, such as the microcontroller ATmegaS128 by ATMEL, that presents a radiation tolerance profile compliant with a LEO orbit. In the same paper, an extensive report on validating SEE caused by protons on COTS like the SPC56EL70L5 by STM, is also reported. The device tested with protons is the same studied in this paper.

Other studies on the space environment and its effect on electronic components are reported in [2], where some solutions to mitigate the damages are presented. The most relevant of these are: the implementation of two equivalent systems that work in redundancy, in which a reset is performed if there is a disagreement between them; the application of a watchdog timer, where the system is reset if a pre-settled action is not registered in a specific time interval.

Our study will focus on the validation of a microcontroller by STMicroelectronics, namely SPC56EL70L5. Two tests will be presented: one performed with heavy-ions, the other with atmospheric-like neutrons. Protons, neutrons and heavy ions are present in space, particularly in Low Earth Orbit (LEO), where many satellites and cube-sat constellation will be stationed in the near future.

The paper is structured as follows: section II describes the tested device; the facilities at which the tests were performed are illustrated in section III; section IV describes the experimental methodology; the data analysis is reported in section V and finally section VI summarizes the conclusions.

## II. THE SPC56EL70L5 MICROCONTROLLER

The Device Under Test (DUT) is the SPC56EL70L5, a 32-bit Power Architecture® microcontroller, manufactured by STMicroelectronics, and developed for automotive chassis and safety applications that require a high Safety Integrity Level (SIL). The device is designed with the possibility to configure it as a dual lock-step and, thanks to this configuration, it can achieve IEC61508 SIL3 and ISO26262 ASILD integrity levels.

A detailed block diagram of the device is shown in Fig. 1.

The dual lock-step is provided with dual redundancy for the essential components of the device, represented by the cyan blocks in Fig. 1; this redundancy allows the device to reach the targeted SIL with minimal additional software and hardware.

The critical components receive the same initialization values and execute the same operations; the outputs of these (displayed by the red blocks in Fig. 1) are compared to check for errors. An error is reported if there is a discrepancy of the outputs.

The chip is a CMOS device contained in a LQFP144 package of dimension 20mm×20mm×1.4mm, and has 144 pins useful for alimentation, I/O peripherals, power management, ADCs (illustrated by the yellow blocks in Fig. 1).

Each microcontroller is provided with the following memory:
- 2MB of FLASH memory;
- 192 kB of SRAM memory.

The device operates at a frequency up to 120MHz and it is optimized for low power consumption while maintaining high-performance processing power.

A full explanation of the chip architecture is available in [3].

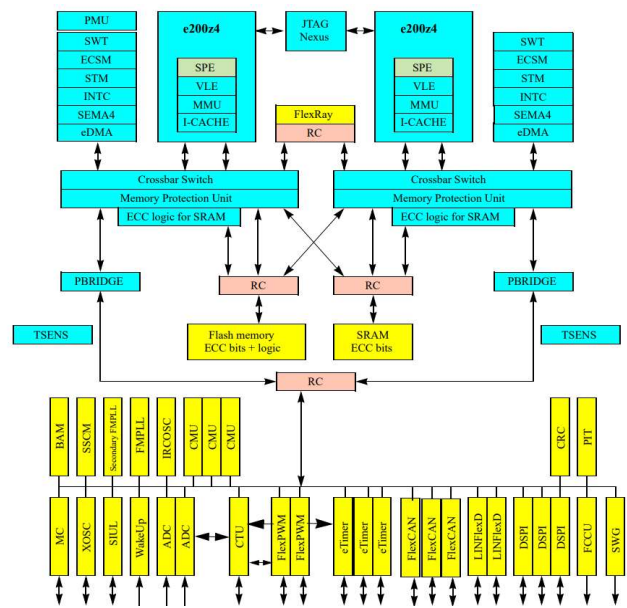

Fig. 1. Block diagram of the SPC56EL70L5 microcontroller under test. Source: [3].

## III. LABORATORIES

The devices were tested under two different kinds of radiation: heavy-ions (Kr) and neutrons. To do so, the DUTs were used, respectively, at *Laboratori Nazionali del Sud* (LNS), a facility of the *Istituto Nazionale di Fisica Nucleare* (INFN), in Catania, Italy, and at the ISIS Neutron and Muon Source at the *Rutherford Appleton Laboratory* (RAL), in UK.

### A. Laboratori Nazionali del Sud (INFN-LNS)

The Laboratori Nazionali del Sud (LNS) of the Istituto Nazionale di Fisica Nucleare (INFN) in Catania, Italy, hosted the ion measurements, performed with the 0° beam line. The particles are accelerated in the vacuum with the Superconducting Cyclotron (CS) up to 80MeV/nucleon. Just before exiting in the air, through a 50μm layer of Kapton, the beam is spread using a 15μm foil of Tantalum.

To ensure that the energy deposition happens inside the device and to carefully evaluate the energy deposition inside the silicon, the Integrated Circuits (ICs) were decapped before the irradiation. Moreover, a 20mm diameter beam was used to ensure that the whole chip is irradiated homogeneously.

For the purpose of our measurements two different ions were used: $^{84}Kr$ accelerated to 1678.24 MeV and $^{78}Kr$ accelerated to 780 MeV. The two ions have a LET of 45MeV·cm²·mg⁻¹ and of 34MeV·cm²·mg⁻¹ respectively.

### B. ISIS Neutron and Muon Source

Neutron measurements were performed at ISIS Neutron and Muon Source at Rutherford Appleton Laboratories (RAL) in Oxfordshire, UK [4,5].

The ISIS spallation source uses a synchrotron (163m circumference) to accelerate protons up to 800MeV; the proton beam is extracted and directed on a tungsten target to produce neutrons. There are two target stations used for different experiments. For our measurements we used target station 2 where there is a dedicated beam line for irradiating microelectronics with atmospheric-like neutrons, the ChipIr instrument [6], that can produce neutrons with energies $E_n > 10MeV$ and with a flux of $5 \cdot 10^6$ neutrons·cm⁻²·s⁻¹ (see e.g., Refs. [7-9]). Through measurements, it has been shown the beam has a 70mm×70mm square shaped uniform footprint. The energy spectrum of the neutron is shown in Fig. 2.

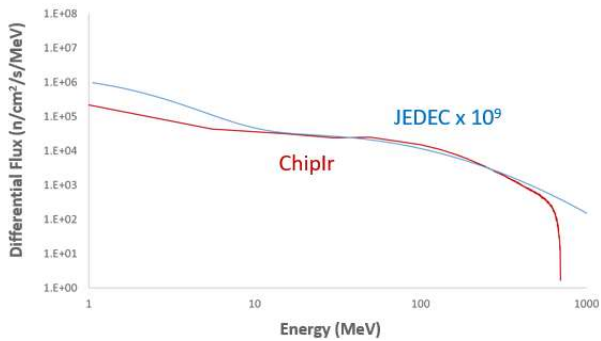

Fig. 2. Neutron energy spectrum (continuous red line), evaluation of 2019, compared to the atmospheric one (blue line). Source: [10, 11].

## IV. METHODOLOGY AND DATA ACQUISITION

To perform the tests, we defined the two types of board used:
- Control board: a remote board, not irradiated, on which the electronics for controlling and powering the microchips based on an FPGA device are mounted; this board is shown in Fig. 4.
- Test board: a printed circuit on which the DUT is mounted; this board is equipped with the useful connector to bring the microchips pins to the control board and a BJT transistor for internal voltage regulation; this board is shown in Fig. 5 (front) and in Fig. 6 (back). After each irradiation, the test board was replaced with a new one.

The two boards communicate with each other via IDC connections. The control board has a USB 2.0 connection to communicate with a pc for controlling the acquisition software and for transferring the acquired data. The main panel of the acquisition software is shown in Fig. 3.

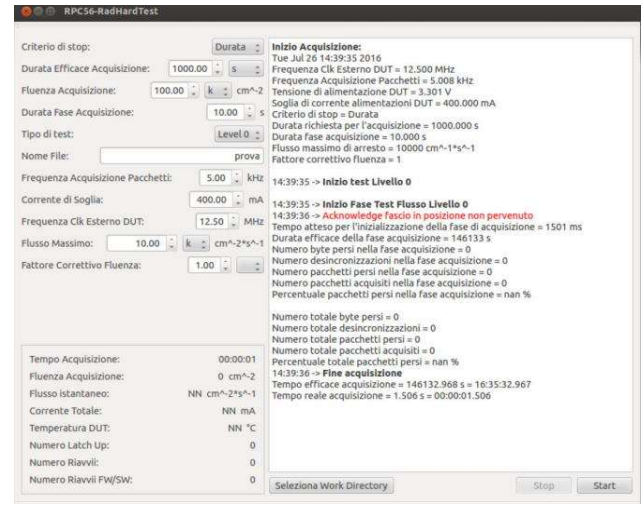

Fig. 3. Main panel of the control software. It is possible to set the acquisition time, the current threshold for latch-up, the clock frequency and parameters for the flux. The software monitors and shows all the set parameters and gives feedback on the number of resets.

On the control board there are 14 voltage supply channels, each of these can be monitored individually and, via shunt resistances or Hall-effect sensors, the corresponding currents are acquired. These currents are listed in Table I.

On the FPGA there is a control firmware that checks the sum of the measured currents; if this sum is greater than a threshold, the power supply is turned off to protect the DUT, this defines a Single Event Latch-up (SEL). After a SEL occurs the power supply is not available for 2s.

TABLE I
MONITORED CURRENTS WITH PINS

| Pin Number | Current | Current Description |
|---|---|---|
| 6 | I_IO | I/O pin |
| 16 | I_REG_0 | PMU regulator |
| 21 | I_IO | I/O pin |
| 27 | I_OSC | Internal oscillators |
| 50 | I_ADDR0 | ADC0 reference |
| 56 | I_ADDR1 | ADC1 reference |
| 58 | I_ADV | Integrated ADC power supply |

| 72 | I_PMU | Power Management Unit (PMU) |
|---|---|---|
| 91 | I_IO | I/O pin |
| 95 | I_REG_1 | PMU regulator |
| 97 | I_FLA | 2MB Flash memory |
| 126 | I_IO | I/O pin |
| 130 | I_REG_2 | PMU regulator |
| EXTERNAL | I_Hall | Whole current monitor |

The firmware installed inside the DUT generates a square pulse on pin GPIO 79. This pin is monitored by the FPGA to check if the core is working correctly. When the chip is positioned under the irradiating beam an additional signal is required to ascertain that the CPU is working: an *acknowledge* signal in response to a control signal given by the FPGA. This acknowledge signal is not mandatory, but it is rather an additional check.

If the chip does not receive any waveform on the GPIO for more than 6.7s, a hard reset is performed on the power supply because the execution of the firmware is considered off. This process can be executed as long as desired, usually a 40s cycle is selected, referred to as a GPIO test.

The purpose of the experiment is to study the behavior of the chip under accelerated irradiation. During the irradiation, the currents are monitored in two different phases: GPIO phase and Beam phase. The former is the phase when the DUT is irradiated, the latter is used to monitor the beam flux intensity, since the ion beam flux was not monitored by the LNS facility. In addition, the current absorption is measured while the samples are maintained at 80°C.

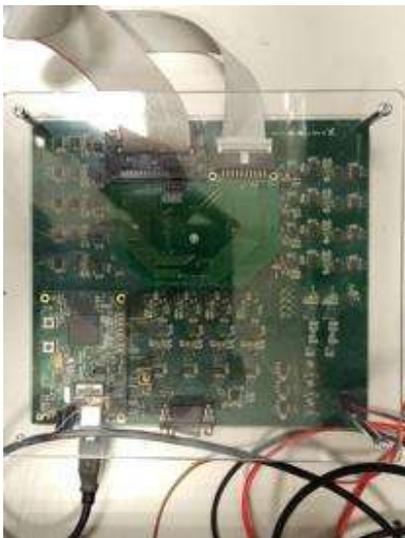

Fig. 4. Control board. The IDC connectors on top bring the power supply to the test board. The USB connector at the bottom gives access to the acquisition software. This board is put outside the irradiation rooms.

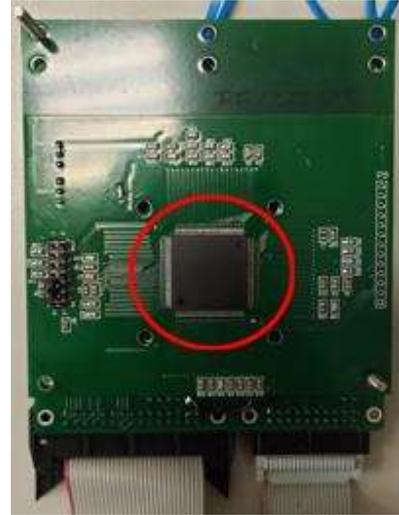

Fig. 5. Test board, front side. The DUT is highlighted by the red circle and is powered up by the control board via IDC connectors

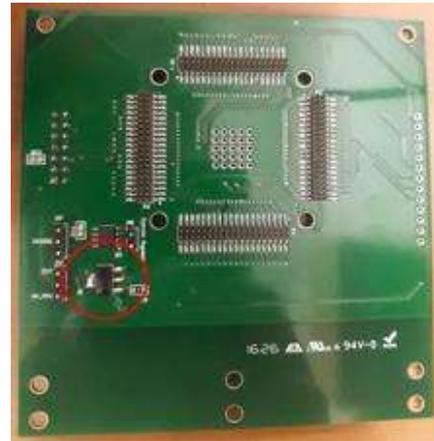

Fig. 6. Test board, back side. The BJT transistor for internal voltage regulation is highlighted by the red circle.

## V. DATA ANALYSIS

Both experiments have been performed following the descriptions made in section IV.

### A. Heavy-ions beam test results

A total of eight samples were irradiated with a heavy-ion beam: five with $^{84}$Kr of energy 1678.24MeV and LET of 45MeV·cm²·mg⁻¹; three with and LET of 34MeV·cm²·mg⁻¹. The values for the LET were obtained with a simulation using the software SRIM2013 [12].

To avoid attenuation on the surface of the LQFP144 package the device was decapped as shown in Fig. 7. The decapping was performed by STMicroelectronics using a mixture 3:1 of nitric acid ($HNO_3$) and sulphuric acid ($H_2SO_4$), at room temperature, dosed with an automatic dispenser.

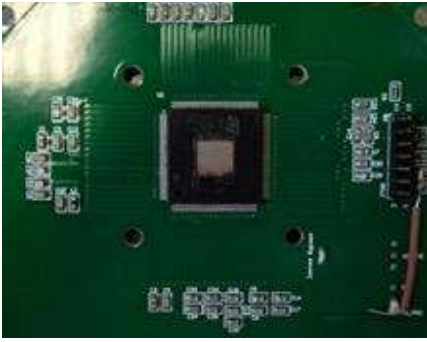

Fig. 7. Close-up of the decaped chip. Part of the LQFP package was removed with the acid solution.

For these tests a mechanical device was implemented to control and measure the ion flux. The DUTs were mounted on a mechanical translator that moves the microcontrollers under the beam or away from it. In line with the beam, behind the device, a scintillator crystal coupled to a photomultiplier was positioned. The irradiation of the DUT lasts for about 10 minutes, then the DUT is moved away to let the ions impinge on the scintillator that is irradiated for about 40s. With this method an ion counting was performed. A schematic of the procedure is shown in Fig. 8. In the figure the movement of the translator is represented by the black double-headed arrow.

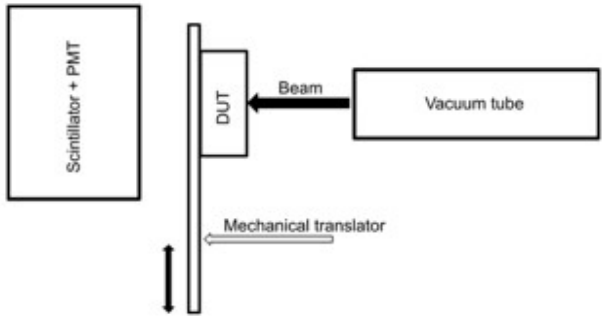

Fig. 8. Schematic of the irradiation and ion counting. When the translator moves down the beam hits the scintillator and the ion counting starts.

This mechanism represents the two phases of the test described in section IV: GPIO phase and Beam phase. At the time of the experiment it was not possible to directly control the low fluxes of the ion beam, so it was crucial to implement a system that allowed the measurement and monitor of the irradiation received by the DUT. For both the irradiation the latch-up threshold is set to 1A.

During both irradiations the temperature of the device is stabilized to $(79.5\pm1.0)°C$ using a Peltier cell equipped with a control loop, composed of a *Proportional Integral Derivative* (PID) controller, to monitor the hysteresis. At this temperature the absorbed current, before irradiation, has a mean value of $(50.0\pm0.5)$mA.

In Table II are reported the results of the test with the ion $^{84}$Kr; in Table III the results with the ion $^{78}$Kr. The absorbed current reported is the one after irradiation.

The flux has an uncertainty given by natural radioactivity which is equal to $(20\pm5)$ ions·s$^{-1}$·cm$^{-2}$.

The effective fluence is obtained by multiplying the mean flux by the irradiation time. The uncertainty on this quantity is given by the integral over time of the natural radioactivity fluence which is equal to $(20\pm5)\cdot10^4$ ions·cm$^{-2}$, for an interval of $10^4$s (ion $^{84}$Kr), and $(20\pm5)\cdot10^3$ ions·cm$^{-2}$, for an interval of $10^3$s (ion $^{78}$Kr).

<div align="center">

TABLE II
$^{84}$Kr IRRADIATION RESULTS

</div>

| | | HEAVY-ION: $^{84}$Kr | | |
|---|---|---|---|---|
| Sample | Fluence ($10^7$ions·cm$^{-2}$) | Irradiation time (s) | Absorbed current (mA) | Test passed |
| ST01 | 1.01 | 6027 | 50.0±0.5 | Yes |
| ST02 | 1.21 | 6321 | 30.0±0.5 | No |
| ST03 | 1.33 | 8205 | 50.0±0.5 | Yes |
| ST05 | 1.27 | 10790 | 50.0±0.5 | Yes |
| ST06 | 0.21 | 3592 | 50.0±0.5 | Yes |

<div align="center">

TABLE III
$^{78}$Kr IRRADIATION RESULTS

</div>

| | | HEAVY-ION: $^{78}$Kr | | |
|---|---|---|---|---|
| Sample | Fluence ($10^7$ions·cm$^{-2}$) | Irradiation time (s) | Absorbed current (mA) | Test passed |
| ST04 | 1.14 | 13813 | 50.0±0.5 | Yes |
| ST07 | 0.23 | 2706 | 50.0±0.5 | Yes |
| ST08 | 0.026 | 900 | 50.0±0.5 | Yes |

A test is considered passed, i.e. the chips survived the irradiation, if the current absorbed and the contents of the EEPROM matches before and after the irradiation.

Of the eight microcontrollers irradiated only one, ST02, did not pass the test; the absorbed current for this was $(30.0\pm0.5)$mA.

For these samples we computed the absorbed dose. Since this is proportional to the fluence and the energy loss rate in a material [13], we computed it by using (1):

$$D = 1.602 \cdot 10^{-7} \cdot \phi \cdot LET, \qquad (1)$$

where $D$ is the dose (expressed in Gy=J·kg$^{-1}$), $\phi$ is the fluence (expressed in #particles·cm$^{-2}$) and $LET$ is the Linear Energy Transfer (expressed in MeV·cm$^2$·mg$^{-1}$). The numerical factor is a conversion that changes MeV to J and mg to kg. The value obtained is from 15.14Gy to 95.9Gy with the ion $^{84}$Kr, and from 1.4Gy to 62Gy with $^{78}$Kr.

For each sample we also evaluated the number of SEL, the number of firmware (FW) blocks and the SEL-FW cross-section defined as (2):

$$\sigma = \frac{\#SEL + \#FW\,block}{fluence}. \qquad (2)$$

The results of this computation are reported in Table IV. From this table it is clear that the only events observed are due to FW blocks, and not to SEL occurred.

From Tables II, III and IV we can conclude that the DUT has a good resistance to heavy-ions fluences up to $10^7$ ions·cm$^{-2}$, since only one of eight did not pass the test. However, a hot redundancy is needed to ensure the service of the data acquisition because we observed FW interruption during the irradiation.

TABLE IV
NUMBER OF SEL AND FW BLOCK WITH CROS-SSECTION

| Sample | Fluence ($10^7$ions·cm$^{-2}$) | #SEL | #FW Block | $\sigma$(cm$^2$) |
|--------|-------------|------|-----------|---------|
| ST01 | 1.01 | 0 | 816 | 8.08e-5 |
| ST02 | 1.21 | 0 | 802 | 6.63e-5 |
| ST03 | 1.33 | 0 | 56 | 4.21e-6 |
| ST04 | 1.14 | 0 | 479 | 4.20e-5 |
| ST05 | 1.27 | 0 | 1248 | 9.83e-5 |
| ST06 | 0.21 | 0 | 130 | 6.19e-5 |
| ST07 | 0.23 | 0 | 120 | 5.22e-5 |
| ST08 | 0.026 | 0 | 15 | 5.77e-5 |

The ion environment in a real orbit has much lower flux/fluence than our experiment [14]. To compare our result with the real environment we have computed the FW block rate for our experiment and the Geosynchronous Equatorial Orbit (GEO) and LEO environments. The results of this computation are shown in Table V, where we have taken as example samples ST01 and ST04.

TABLE V
FLUX AND EVENT RATES FOR THE EXPERIMENT, LEO AND GEO

| LET (MeV·cm2·mg-1) | Environment | $\sigma$(cm$^2$) | Flux (cm$^{-2}$·s$^{-1}$) | FW Blocks rate (s$^{-1}$) |
|--------|-------------|---------|---------|---------|
| | EXPERIMENT (ST01) | | 1.68e3 | 1.35e-1 |
| 45 | | 8.08e-5 | | |
| | LEO | | 2.31e-8 | 1.87e-12 |
| | GEO | | 6.37e-9 | 5.14e-13 |
| | EXPERIMENT (ST04) | | 8.25e2 | 3.47e-2 |
| 34 | LEO | | 8.10e-8 | 3.40e-12 |
| | GEO | 4.20e-5 | 2.08e-8 | 8.75e-13 |

In the case of the experiment performed, there are more FW blocks than the ones expected in the real LEO and GEO environment; this is due to the higher fluxes used to accelerate the experiment. Considering ST01, as an example, this sample registers a period between FW blocks of about 7.41s, compared to the $5.35 \cdot 10^{11}$s and $1.94 \cdot 10^{12}$s for the LEO and GEO respectively. For a three years mission these values correspond to a number of FW blocks as follows: $1.76 \cdot 10^{-4}$ for LEO and $4.85 \cdot 10^{-5}$ for GEO.

### B. Neutron beam test results

The neutron beam experiment lasted 65 hours divided into four days, from the 27th of April 2021 to the 30th of April 2021. Seven samples have been tested with neutron fluences on a single sample with a fluence up to $10^{11}$neutrons·cm$^{-2}$.

In order to keep the same configurations, the DUT and the acquisition software are the same as those used in the heavy-ions tests but, in this case, there is not the real necessity to have the GPIO phase and the Beam phase since the fluence of the beam is directly measured by the hosting facility.

To begin with, the samples are tested in the experimental setup, with the neutron beam off, for a couple of minutes to collect some references. In a second time, the samples get irradiated and tested for several hours. They get finally tested with the beam off again, to check the state and the behavior of the chips during ten minutes after the irradiation.

The total fluence irradiated during this period can be seen in Fig. 9. The grey area represents the effective time under beam for each sample, labeled in the bottom left corner of the area. The points indicate the breaking time of the damaged chips.

The exposure time of each sample is reported in Table VI alongside the fluence the sample were exposed to, both total and before breaking, and a tag reporting if the chip has broken during the irradiation or not.

For our measurement, we defined a broken chip as one that resets continuously -every 7s- during the GPIO phase of the test because, as stated in section IV, this defines the hard reset and in this situation the chip cannot recover the firmware.

We monitored the currents registered during the irradiation to check for this pattern of resets, an example of which is shown in Fig. 10. The vertical axis is the current, registered on the ADC power supply, during irradiation time, expressed in mA; on the horizontal axis the time from the start of acquisition, expressed in seconds. Before the breaking point (represented by the dashed line) the device underwent resets, but these were not continuous and every 7s; instead, after the break, the current dropped of 1.5mA and the microcontroller started resetting continuously. Every bunch of reset after break represents the GPIO phase of acquisition and every bunch is 40s away from the following, as selected via the acquisition software.

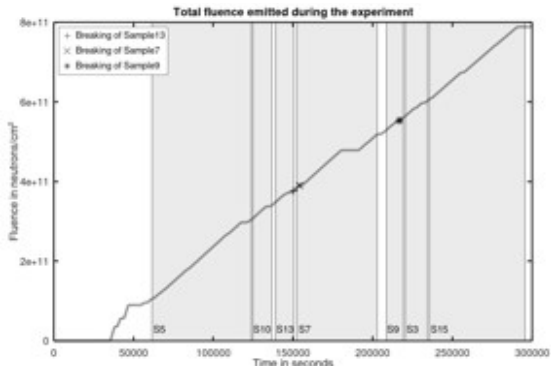

Fig. 9. Total fluence registered by RAL. Each grey area represents the irradiation time of a chip, whereas the points the time of break for samples: 7 (✕), 9 (*) and 13 (+).

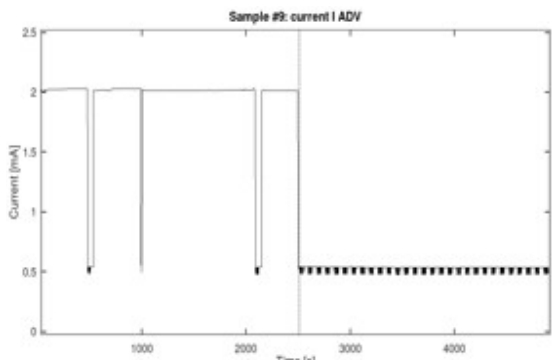

Fig. 10. Example of the current acquired by a broken chip (sample 9). The vertical dotted line represents the breaking point, on the right side of this we can see the pattern of resets typical of a broken chip. The current shown is the one from the ADC voltage supply during irradiation time.

The number and frequency of resets measured in each sample are the main parameters to distinguish a chip which broke during the irradiation from a chip which did not. Among the seven we tested, three of them (samples 7, 9 and 13) behaved like broken ones after a certain amount of time.

TABLE VI
NEUTRON IRRADIATION RESULTS

| Sample | Total fluence (neutrons·cm²) | Fluence before breaking (neutrons·cm⁻²) | Irradiation time (s) | Broken |
|--------|------------------------------|------------------------------------------|----------------------|--------|
| S3 | 3.78e10 | -- | 13885 | No |
| S5 | 1.98e11 | -- | 62220 | No |
| S7 | 1.34e11 | 5.89e9 | 50365 | Yes |
| S9 | 3.06e10 | 2.31e10 | 10543 | Yes |
| S10 | 3.27e10 | -- | 11938 | No |
| S13 | 2.86e10 | 2.86e10 [a] | 11361 | Yes |
| S15 | 1.82e11 | -- | 59886 | No |

[a] During the whole post-irradiation test, S13 acted broken, whereas it did not during the irradiation test. We assumed this chip broke at the end of the acquisition, so the two fluences reported are the same.

To check the status of all of the chips after the irradiation, especially for the broken ones, other tests without the beam were performed two months after the irradiation time. The two month time gap was necessary to allow the neutron-induced activity of the samples to decay and ensure that they were safe to handle.

The non-beam radiation-less tests lasted two hours and were performed connecting the test board to the control board and monitoring the aforementioned 14 currents. For a fine chip we expect no resets, whereas a damaged one should show resets even if in a "no-radiation" environment. Thus, we define a damaged chip as a sample which provides resets during a test without radiation but does not show continuous resets.

After performing radiation-less tests on microcontrollers, S9 was still manifesting the broken behavior (a total of 547 resets) during the two-hour test. However, S7 and S13 did not show this pattern despite the presence of resets (respectively 8 and 30), which should not occur for a fine chip not previously irradiated, thus making these samples at least partially damaged.

To explain this change in the number of resets we supposed that an effect of self-annealing occurred during the two months between the tests. No specific analysis were carried out but, from literature [14], we found some evidence of self-annealing in semiconductors at room temperature that occurred over periods of 45, 60 and 100 days. Despite better results regarding the number of resets, the annealing effect did not make the chips completely recover. Indeed, besides having currents behaving like working samples, the chips do not act like broken ones, displaying a broken pattern. However, unexpected resets are observed, which proves the chips are still partially damaged.

We can establish that the neutron beam irradiation didn't result in SEL behavior of the DUT, taking into account that the used beam has an intensity several orders higher than the natural LEO environment.

The reset pattern of the broken chips can derive either from the DUT in the front of the test board, Fig. 4, or from the BJT transistor on the back, Fig. 5. As a consequence, it is possible that only one of these components is broken, or both of them are. To determine which of the two devices is the reason for this behavior, a solution is to change the transistor with a known-fine one. This modification was made on S7, S9, and S13.

Once again, a two-hours radiation-less test was made for these three samples, with the transistors changed, and respectively 32, 16, and 12 resets were counted. The number of resets for S9 decreased significantly, and the behavior previously described was no longer observed. The number of resets for S7 and S13 is still in the same range as previously mentioned.

In Table VII, the number of resets for each phase of the analysis are summarized: irradiation, radiation-less and radiation-less after changing the transistor (this last one only for samples 7, 9 and 13).

TABLE VII
SAMPLES RESETS SUMMARY

| Sample | Resets during irradiation | Resets during radiationless test | Resets after 60 days | Resets after changing the transistor |
|--------|---------------------------|----------------------------------|----------------------|--------------------------------------|
| S3 | 20 | 2 | 0 | -- |
| S5 | 29 | 20 | 0 | -- |
| S7 | 3782 | 30 | 8 | 32 |
| S9 | 194 | 18 | 547 | 16 |
| S10 | 147 | 4 | 0 | -- |
| S13 | 181 | 62 | 30 | 12 |
| S15 | 24 | | 0 | -- |

These values mean the microcontroller is broken for each of these three samples. Besides, during the irradiation tests, the transistors and the microcontroller broke, but the two-month waiting made an annealing process occur for the transistors of S7 and S13.

The results from Tables VI and VII allow us to say that the DUT has a moderate resistance to neutron fluences up to $10^{11}$ neutrons·cm$^{-2}$, since three out seven appear to be damaged. Changing the BJT helped lowering the number of resets for S9 and S13, whereas S7 increased the number of resets. In the case of S7, a functioning BJT could correctly polarize the DUT and transmit the resets as it should. Thus, the right number of resets occur and are observed.

## VI. CONCLUSIONS

The purpose of this work was to study the resistance and the behavior of the SPC56EL70L5 microcontroller under irradiation of two different particle beams. In order to do it, seven chips were exposed to atmospheric-like neutrons, and eight to heavy-ions. Such study was performed looking at the number of resets to the chip, in the case of neutrons. For heavy-ions, a comparison between the currents before and after irradiation was done. We kept the same configuration for both experiments on purpose so as to keep a matching in the bugs, the functioning and behaviors.

We observed that:

● The microcontroller shows sufficient resistance to neutron beam fluxes up to $10^{11}$ neutrons·cm$^{-2}$. Of the seven samples tested three (S7, S9, S13) showed a broken pattern immediately after irradiation but, after two months, only one of these (S9) maintained the broken pattern; S7 and S13 showed regular currents and a reduced number of resets. A possible explanation is attributed to an annealing effect that took place during the two months between the irradiation and the radiation-less tests; but no further analysis was conducted. As the origin of resets could be the chip as much as the transistor of the board, we changed the latter. After this change, the current didn't show the broken pattern anymore, but there were still resets, implying the chips were damaged.

● The DUTs show a good resistance to heavy-ion beam fluxes up to $10^{7}$ ions·cm$^{-2}$. Only one (ST02) out of eight samples did not pass the test presenting an absorbed current after irradiation lowered by 40%. All the observed events are due to firmware block, no SEL occurred.

ACKNOWLEDGMENT

The authors gratefully acknowledge Regione Lazio (IR approved by Giunta Regionale Determinazione n. G10795 070/8/2019, published by BURL n. 69 August 27/08/2019) and University of Rome "Tor Vergata" for the financial supports to ISIS@MACH Regional Project.

The financial support from the University of Rome "Tor Vergata (UNITOV) within the Amendment to MOU between UNITOV-STFC regarding the mutual exploitation of ISIS and ISIS@MACH [originated from the common interest of the parties in the exploitation of ISIS@MACH, a Research Infrastructure funded by Region Lazio] is gratefully acknowledged.

The financial support from the Consiglio Nazionale delle Ricerche within CNR-STFC Grant Agreement [No. 2014-2020 (N 3420)], concerning collaboration in scientific research at the ISIS Neutron and Muon Source (UK) of Science and Technology Facilities Council (STFC), is gratefully acknowledged.

REFERENCES

[1] G. Furano, *et al.* (2016, september). A novel method for SEE validation of complex SoCs using low-energy proton beams. Presented at 2016 IEEE International Symposium on Defect and Fault Tolerance in VLSI and Nanotechnology Systems (DFT). [Online]. Available: https://www.researchgate.net/publication/309496330_A_novel_method_for_SEE_validation_of_complex_SoCs_using_Low-Energy_Proton_beams. Accessed on: September 07, 2021.

[2] Md M. Rahman, D. Shankar, S. Santra (2017, september). Analysis of radiation environment and its effect on spacecraft in different orbits. Presented at International Astronautical Congress (IAC2017), Adelaide, AUS. [Online]. Available: https://www.researchgate.net/publication/320765337_Analysis_of_Radiation_Environment_and_its_Effect_on_Spacecraft_in_Different_Orbits. Accessed on: September 08, 2021.

[3] RM0342 - SPC56xL70xx Reference manual, STMicroelectronics, October 2013.

[4] UK Research and Innovation, Science and Technology Facilities Council, ISIS Neutron and Muon Source. *How ISIS works - in depth.* URL: https://www.isis.stfc.ac.uk/Pages/How-ISIS-works--in-depth.aspx. Accessed on: September 15, 2021.

[5] UK Research and Innovation, Science and Technology Facilities Council, ISIS Neutron and Muon Source. *ChipIr technical information.* URL: https://www.isis.stfc.ac.uk/Pages/Chipir-technical-information.aspx. Accessed on: September 15, 2021.

[6] C. Andreani, A. Pietropaolo, A. Salsano, G. Gorini, M. Tardocchi, A. Paccagnella, S. Gerardin, C.D. Frost, S. Ansell, S.P. Platt, "Facility for fast neutron irradiation tests of electronics at the ISIS spallation neutron source" (2008, March), Applied Physics Letters, 92, 11, 114101

[7] A. Haran, N.M. Yitzhak, E. Mazal-Tov, E. Keren, D. David, N. Refaeli, E. Preziosi, R. Senesi, C. Cazzaniga, C.D. Frost, T. Hadas, U. Zangi, C. Andreani, "Ultralow Power System-on-Chip SRAM Characterization by Alpha and Neutron Irradiation" (2021, September), IEEE Transactions on Nuclear Science, 68, 11, 2598-2608

[8] M. Bagatin, S. Gerardin, A. Paccagnella, C. Andreani, G. Gorini, C.D. Frost, "Temperature dependence of neutron-induced soft errors in SRAMs" (2012, January), Microelectronics Reliability, 52, 1, 289-293

[9] UK Research and Innovation, Science and Technology Facilities Council, ISIS Neutron and Muon Source. *ChipIR*. URL: https://www.isis.stfc.ac.uk/Pages/ChipIR.aspx. Accessed on: October 12, 2021.

[10] Chiesa, Davide, et al. "Measurement of the neutron flux at spallation sources using multi-foil activation." *Nuclear Instruments and Methods in Physics Research Section A: Accelerators, Spectrometers, Detectors and Associated Equipment* 902 (2018): 14-24.

[11] Cecchetto, Matteo, et al. "SEE flux and spectral hardness calibration of neutron spallation and mixed-field facilities." *IEEE Transactions on Nuclear Science* 66.7 (2019): 1532-1540.

[12] Ziegler, James F.; Ziegler, M. D.; Biersack, J. P, " SRIM - The stopping and range of ions in matter (2010)". *Nuclear Instruments and Methods in Physics Research Section B, 10.1016/j.nimb.2010.02.091*

[13] SLVK046 Application report - Heavy Ion Orbital Environment Single-Event Effects Estimations, Texas Instruments, May 2020. URL: https://www.ti.com/lit/an/slvk046/slvk046.pdf. Accessed on: October 18, 2021.

[14] V. D. S. Dhaka, *et al.*, "Room-temperature self-annealing of heavy-ion-irradiated InGaAs/GaAs quantum well", IEEE *Electronic Letters*, vol. 41, no. 23, Dec. 2005, 10.1049/el:20053117.